\documentclass{jfm}
\usepackage{graphicx} %%For loading graphic files}

\usepackage{natbib} % bibliography
\usepackage{amsmath}
\usepackage{amsthm}
\usepackage{amsfonts}
\usepackage{amssymb}
\usepackage{mathrsfs}
\usepackage{booktabs}

\title{Instability of particulate pipe flow}
\author{Anthony Rouquier, Alban Poth\'erat and Chris C.T. Pringle}

\begin{document}
\maketitle

\begin{abstract}
We present linear stability analysis for a simple model of particle-laden pipe flow. The model 
consists of a continuum approximation for the particles two-way coupled to the fluid velocity field 
via Stokes drag \citep{saffman62}. We extend previous analysis in a channel \citep{klinkenberg11} 
to allow for the initial distribution 
of particles to be inhomogeneous and in particular consider the effect of allowing the particles 
to be preferentially located around one radius in accordance with experimental observations. 
This simple modification of the problem is enough to alter the stability properties of the flow, 
and in particular can lead to a linear instability at experimentally realistic parameters. The 
results are compared to the experimental work of \citet{matas2004jfm} and are shown to be consistent 
with the reported flow regimes.

\end{abstract}

\section{Introduction}
This paper is concerned with the wide issue of how particles affect the transition to turbulence in a 
pipe flow.
Beside the fundamental interest of this canonical problem, several industrial sectors have seen a 
growing need to accurately measure flow rates or volume fractions in complex fluid mixtures flowing 
through pipes. Examples range from the precise determination of the volume fraction of oil in the 
oil-water-sand-gas mixture that is extracted from offshore wells, to needs in the food processing 
industry \citep{ismail2005fmi}, and flows of molten metal carrying impurities during recycling processes 
\citep{kolesnikov2011mmtb}. Each of these examples requires dedicated flow metering technologies, most 
of which rely on \emph{a priori} knowledge of the nature of the flow inside the pipe or the duct and 
in particular whether it is turbulent or not \citep{wang2014fmi}. 
Though none of these examples could satisfactorily be modelled as a single fluid phase carrying 
one type of particles, the ideal problem of the particulate pipe flow constitutes one of their 
elementary building blocks. As such it is a good starting point from which to infer the basic mechanisms governing the
 transition to turbulence.

Adding particles opens a number of possibilities, associated with 
different physical mechanisms: particles can be buoyant or not, of different sizes and shapes, and also 
mono- or polydisperse. As a first step in studying the transition to 
turbulence in particulate pipe flows, we shall focus on the simpler case of neutrally buoyant, 
monodisperse spherical particles.
Whether the effect of particles on the transition to turbulence in general is a stabilising or 
destabilising one mostly depends on the size and volume fraction of particles.
Early experiments on the transition to turbulence in a pipe highlighted a critical volume fraction of 
particles below which they favoured the transition at a lower Reynolds number. At higher volume fractions
 than this critical value, by contrast, the effect was reversed \citep{matas2003prl}. 
  Recent numerical 
simulations based on accurate modelling of individual solid particles recovered this phenomenology 
\citep{yu2013pf}. 

The non-trivial nature of the influence of particles is further supported by the 
numerical study of individual perturbations introduced in a channel:  
whilst below a critical volume fraction, particles lower the critical energy beyond which perturbations triggered 
the transition to turbulence, the transition takes place longer after the perturbation was introduced in the presence 
of particles if the perturbation took the 
form of streamwise vortices \citep{klinkenberg13}. At high volume fraction, the critical energy was 
increased. Linear stability analysis in the same configuration provide a hint on the origin of this 
non-monotonous effect of volume fraction: they revealed the existence of an optimal stabilisation regime due to a 
maximum in the Stokes drag, when the particle relaxation time (\emph{i.e.} the time for a particle at rest to accelerate 
to the velocity of the surrounding fluid), coincided with the period of the streamwise oscillation \cite{klinkenberg11}.

Single phase pipe flow is governed by the a sole parameter, the nondimensional flow rate or Reynolds number. The 
problem remains linearly stable even at large Reynolds number \citep{meseguer03}, and so the turbulence that 
is observed even at moderate flow rates (above $Re \simeq 2000$) must be initiated by finite amplitude disturbances.
The inclusion of particles complicates this, and could even lead to linear instability.

Adding particles to the pipe flows raises the question of how particles shall be distributed in the 
pipe, at least in some initial state. 
While a homogeneous spatial distribution may first come to mind as the simplest possible, particles in pipe are known 
to aggregate near a specific radius greater than $65\%$ of the pipe radius \citep{segre1962jfm}, that increases slowly with the Reynolds
 number. The underlying mechanism is driven by the radial variations of the lift force experienced by particles rotating 
 in shear \citep{repetti1964nat}. The dependence of the aggregation radius (often called the \emph{Segr\'e-Silberberg 
 radius}) on the Reynolds number can be explained by means of asymptotic theory introducing the particle Reynolds number 
 as the small parameter in the expression of the lift force \citep{schonberg1989jfm,hogg1994jfm,asmolov1999jsme}. 
 
 While this dependence is well recovered in experiments at moderate Reynolds numbers, a second equilibrium position
  appears at a   lower radius \citep{matas2004jfm} for $Re>600$. Although this transition  
coincides with a change in the concavity of the radial profile of the lift force, the detailed mechanisms 
driving this effect remain to be found, and the authors left open the question of whether this 
equilibrium is stable or not.
\citet{han1999jr} note that the main effect of particle 
concentration on this phenomenology is to disperse the particle distribution around the equilibrium 
annulus. However, higher concentrations can also lead to the formation of trains of particles aligned 
with the stream \cite{matas2004pf}. In the context of the transition to turbulence, the natural 
tendency of particles to aggregate around specific radii at different Reynolds numbers raises the 
question of the critical Reynolds at which these annuli of particles break-up and whether this break-up plays any role in the triggering of turbulence.

The variety of phenomena observed in particulate flows illustrates the numerous aspects of its transition 
to turbulence (starting with the difficulty of even distinguishing turbulent fluctuations from particle-induced ones). 
As such our purpose in the context of the pipe flow shall be limited to first step of 
investigating the linear stability of the particulate pipe flow to infinitesimal perturbations. Tackling 
this question requires the choice of a strategy to model particles (see \citet{maxey2017arfm} for a review on current
methods). While the most accurate method consists of modelling particles as individual solids \citep{uhlmann2005jcp},
this approach is the most computationally expensive and may not allow for consideration of a long enough pipe to cover 
long-wave instabilities. Cost-effective alternatives exist based on individual point-particle model that can incorporate 
various levels of complexity (one or two-way interaction, rotation of particles, particle interaction etc...). However, 
in the spirit of simplicity of this first step, we shall follow   the even simpler option of modelling particles as a 
second fluid phase whose interaction with the fluid phase is limited to the drag forces that each phase exerts on the
other \citep{saffman62, klinkenberg11} . Within this framework we address the questions of whether particulate 
pipe flow is stable for
either homogeneous or inhomogeneous distributions of particles; which distributions of particles most adversely effect 
stability; and whether the distributions are realistic in comparison with experiments.
The paper is organised as follow: in section \ref{sec:model}, we shall introduce the model and the assumptions it relies
as well as the numerical methods used. We shall then start by considering the simplest case of a homogeneous particle
distribution in the pipe (section \ref{sec:uniform}), before studying the linear stability of particles normally 
distributed around a radius, paying particular attention on how the standard deviation and the value of this radius 
influence the flow stability (section \ref{sec:nonuniform}). We then compare our findings to the experiments of 
\cite{segre1962jfm} and \cite{matas2004jfm} (section \ref{sec:expts}), where localisation was 
observed before discussing the possible implications of our results for the transition to turbulence 
(section \ref{sec:conc}). 

\section{Model and governing equations}\label{sec:model}
In order to avoid the heavy computational load cost incurred by when accounting for particles as 
individual solids,
% to the description of particles and the shifting boundary conditions between particles and fluid 
we describe the particulate flow using the ``two-fluid'' model first derived by \cite{saffman62}. 
Particles are described as a continuous field rather than as discrete entities with a finite size.
This model takes into account neither effects due to particle-particle interactions such as collisions 
or clustering, nor the deflection of the flow around the particles around particles.
% is not taken into account. 
It is therefore valid for lower concentrations and in the limit where particles are
sufficiently smaller than the characteristic scale of the flow.

We consider the flow of a fluid (density $\rho$, viscosity $\mu$) through a straight pipe with constant circular 
cross-section of radius $r_0$ and driven by a constant pressure gradient. The fluid carries particles 
of radius $a$. To describe the problem we adopt the model proposed by \citet{saffman62} and studied by 
\cite{klinkenberg11} in the context of channel flow. The particles are considered as a continuous field 
with spatially varying number density $N$,
their motion coupled to the fluid solely via Stokes' drag, $6\pi a\mu (\mathbf{u_p}-\mathbf{u})$. 
We take coordinates $(r,\theta,z)$ with respective 
velocities $\mathbf{u}=(u,v,w)$. Where relevant we distinguish those quantities associated with the particles from 
those associated with the fluid by means of a subscript $p$. 
After nondimensionalising by the centreline velocity, $U_0$, the pipe radius 
$r_0$ and the fluid density $\rho_f$ we have the equations
\begin{align}
&\partial_t \mathbf{u}+ (\mathbf{u} \cdot \nabla ) \mathbf{u} = - \nabla \mathbf{p}    +  \frac{1}{Re}\nabla^2 \mathbf{u} + \frac{f}{S Re} (\mathbf{u_p} -\mathbf{u}),& \label{eqn:nonlin1} \\
&\partial_t \mathbf{u_p} + (\mathbf{u_p} \cdot \nabla) \mathbf{u_p}= \frac{1}{S Re} ( \mathbf{u} -\mathbf{u_p} ),& 
 \\
&\partial_t N = - \nabla \cdot (N \mathbf{u_p}),&  \\
&\nabla \cdot \mathbf{u} = 0 .& \label{eqn:nonlin2}
\end{align}
We have non-dimensional Reynolds numbers $Re=U_0r_0/\nu$, dimensionless relaxation time $S=2a^2\rho_p/9r_0^2\rho_f$
and mass concentration $f=m_p/m_f$, the ratio of total mass of particles to total mass of fluid. These 
equations are augmented with an impermeable and no-slip boundary condition for the fluid 
\begin{equation}
\mathbf{u}\vert_{r=1}=0 \label{eqn:bc_f}
\end{equation}
and a no penetration boundary condition for the particles
\begin{equation}
u_p\vert_{r=1}=0. \label{eqn:bc_p}
\end{equation}

The stability of the flow is studied through the addition of a small perturbation to the steady solution 
($\mathbf{U}=\mathbf{U}_p=(1-r^2)\hat{\mathbf{z}}$)

\[ \mathbf{u}=\mathbf{U}+\mathbf{u}' , \; \; \mathbf{u_p} = \mathbf{U} + \mathbf{u_p}', \; \; p = P + p' , \; \;  N = N_0+N'. \]

Linearising equations (\ref{eqn:nonlin1}) - (\ref{eqn:nonlin2}) around this base state and dropping the primes gives
\begin{align}
&\partial_t \mathbf{u} +\mathbf{U} \cdot \nabla \mathbf{u}  +\mathbf{u} \cdot \nabla \mathbf{U} = -\nabla \mathbf{p}  + \frac{1}{Re} \nabla^2 \mathbf{u}  + \frac{f}{S Re } (\mathbf{u_p} -\mathbf{u}) , \label{eqn:lin1} \\
&\partial_t \mathbf{u_p} +\mathbf{u_p}\cdot \nabla \mathbf{U} + \mathbf{U} \cdot \nabla \mathbf{u_p} =  \frac{1}{S Re} ( \mathbf{u} - \mathbf{u_p} ),  \\
&\partial_t N = - N_0 \nabla \cdot \mathbf{u_p} - \mathbf{u_p} \cdot \nabla N_0 - \mathbf{U} \cdot \nabla N, \label{eqn:linN} \\
&\nabla \cdot \mathbf{u} = 0.\label{eqn:lin2}
\end{align}
The boundary conditions for the perturbation are the same as for the full flow.
\subsection{Linear stability}
Given the streamwise and azimuthal invariance of the problem, we consider perturbations of the form 
\begin{equation}
g(r,\theta,z,t)=\sum_{n=0}^{N} g_nT_n(r)\exp\{i((\alpha z+m\theta-\omega t)\}, \label{eqn:form}
\end{equation}
where $T_n$ is the $n^{th}$ Chebyshev polynomial and $g$ is any of the fields of interest, with corresponding 
coefficients $g_n$ in the expansion. Substituting this into 
equations (\ref{eqn:lin1})-(\ref{eqn:lin2}) and boundary conditions (\ref{eqn:bc_f}) and (\ref{eqn:bc_p}) leads, 
after collocation, to a generalised eigenvalue problem
\begin{equation}
\label{code}
i \omega \mathbf{A} \mathbf{\phi}  = \mathbf{B}  \mathbf{\phi}
\end{equation}
which can be solved using LAPACK.

The code was validated for the case of non-particulate pipe flow against \cite{meseguer03} and for 
particulate flow in a channel against \cite{klinkenberg11}. With 100 Chebychev polynomials, the relative error 
remains below
$10^{-7}$ for the various tested combinations of values of $\alpha\leq1$, $m\leq1$ and $Re<10^4$. With the addition of 
particles, the precision dropped to $10^{-6}$ for $S=10^{-3}$ and down to $10^{-5}$ for $S=0.1$, with as many as 200 
polynomials. 

To further test all cases, we created a linear simulation code based on the non-particulate DNS 
code of \citet{openpipeflow}. 
This code uses Fourier-modes in the axial and azimuthal directions, and finite differences radially. This allowed us to 
check the leading eigenvalue of each different Fourier mode for any given configuration. Accuracy between the 
methods was confirmed to be within 1\%.

\section{Uniform particle distribution}\label{sec:uniform}
Previous work on this model \citep{klinkenberg11} has made the assumption that the initial distribution of 
particles is uniform throughout 
the domain. This simplifies the governing equations as $\nabla N_0=0$, removing terms and decoupling 
equation (\ref{eqn:linN}) from the rest of the problem.

\subsection{Modified eigenvalue spectrum}
\begin{figure}
\centering
\includegraphics[width=0.9\textwidth]{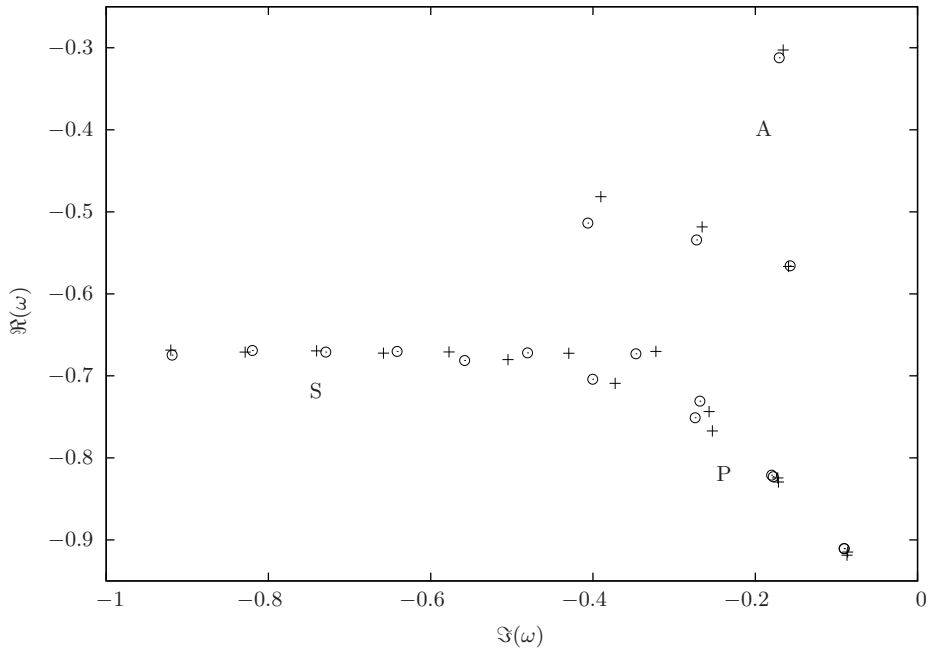}
  \caption{Eigenvalue spectra for the generalised eigenvalue problem (\ref{code}) for \(Re = 1000\), \(S = 10^{-3}\), 
  \(\alpha = 1 \), \(f= 0.1 \). The classical single phase eigenvalues are marked `$+$' while the eigenvalues 
  for the particulate flow are marked `$\circ$'. The three branches of the eigenspectrum are labelled $A$, $P$ and $S$ 
  in accordance with the notation of \citet{mack76}}
\label{fig:u_spec}
\end{figure}

The addition of uniformly distributed particles does not lead to large changes in the stability problem of pipe flow. 
Figure \ref{fig:u_spec} compares the eigenvalue spectra of the particulate and non-particulate problems for one 
typical case. The overall shape of the spectrum is qualitatively unchanged, maintaining three branches, the location 
of the leading eigenvalue being at the tip of the `P'-branch \citep{mack76} . 

We quantify the change in the eigenvalue spectrum by tracking the normalised growth rate
\begin{equation}
\lambda'_p(Re, \alpha,m, f, S) = \frac{\Im\{\omega_p(Re, \alpha, m,f, S)\}}{\Im \{ \omega_f(Re, \alpha,m)\}},
\end{equation}
where $\omega_p$ and $\omega_f$ are the leading eigenvalues in the particulate and non-particulate problems. 
From the definition (\ref{eqn:form}) the growth rate is the imaginary part of the eigenvalue. As the 
pure-fluid problem is linearly stable (meaning $\Im \{\omega_f\}$ is always negative), $\lambda_p'>1$ is indicative of the 
particles stabilising the flow while $\lambda_p'<1$ corresponds to them destabilising the flow. The critical value
$\lambda_p'=0$ would indicate the particulate problem crossing the neutral stability threshold, however this was 
never observed for any parameter combination with a uniform distribution of particles.

\begin{figure}
\centering
\includegraphics[width=0.9\textwidth]{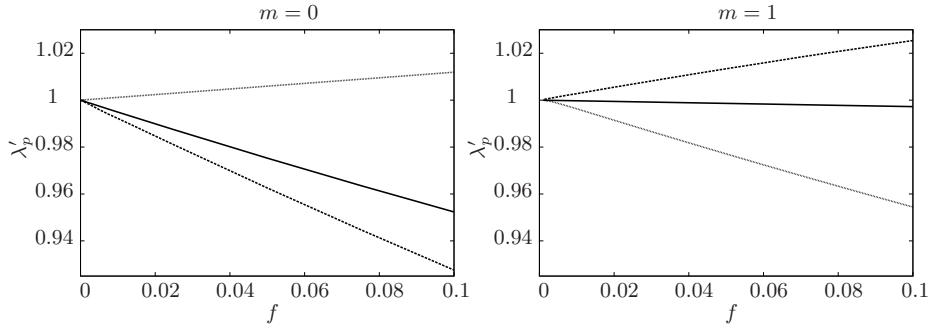}
\caption{Normalised growth rate, $\lambda_p'$, as a function of $f$ for $S=10^{-3}$ (line), $S=10^{-2}$ (dots), 
$S=10^{-1}$ (dashed) with $Re=1000$, $\alpha=1$. In all cases examined, $\lambda_p'$ is very close to being 
proportional to  $f$, suggesting that this parameter simply serves to amplify the underlying result linearly.
\label{fig:u_omega_f}}
\end{figure}

Parameter space is simplified by the observation that the role of $f$, the concentration,  seems to be secondary. 
Figure \ref{fig:u_omega_f} shows $\lambda'_p$ as a function of $f$ and in all cases the concentration serves 
simply to amplify the underlying result almost linearly. Consequently, in the analysis of the uniform particle 
distribution  problem we fix $f=0.01$ in the knowledge that trends could be exacerbated further by increasing the 
quantity.

\subsection{Influence of Stokes number}
Stokes number reflects the size of the particles. It is most easily understood in terms of its limiting 
values. In the ballistic limit, $S\rightarrow\infty$ the large particles become independent of the flow. 
In the other extreme, $S\rightarrow 0$, the particles are passive tracers. In neither case do the particles 
unduly influence the flow. In the former, they fully decouple and one recovers the pure fluid results. In 
the latter case, the particles act as one with the fluid, only changing the effective density of the 
total suspension. This rescales the effective Reynolds number as $Re'=(1+f)Re$ \citep{klinkenberg11}. 

\begin{figure}
\centering
\includegraphics[width=0.95\textwidth]{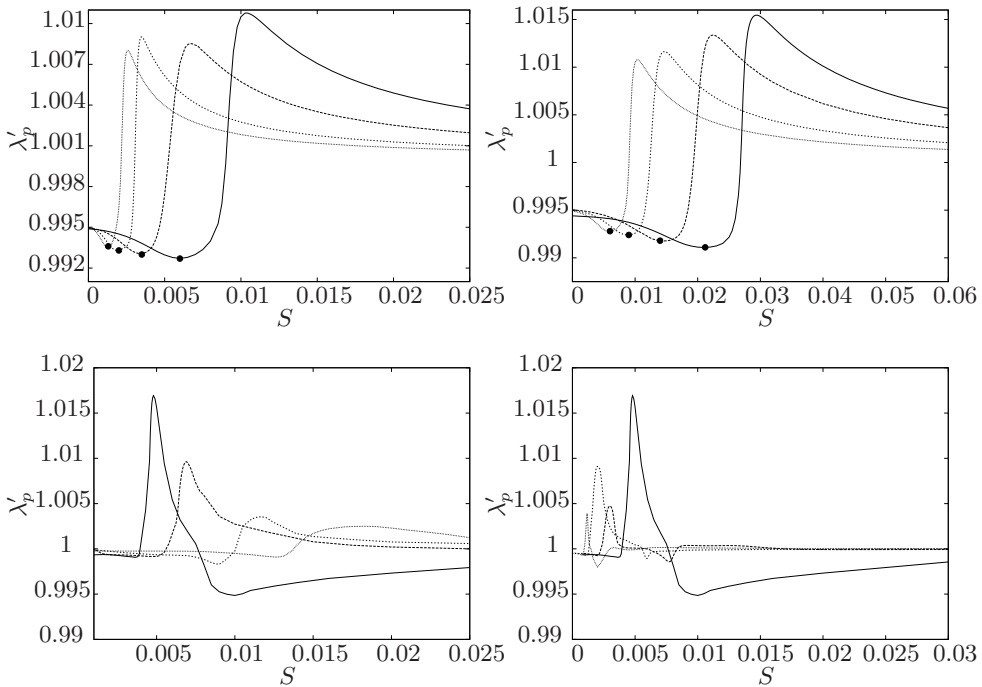}
\caption{Normalised leading growthrate, $\lambda_p'$, as a function of $S$ while keeping $f=0.01$. The top row is $m=0$, 
the bottom is $m=1$. In all cases we 
see that for low Stokes number we recover the result of a single fluid, albeit of larger effective density. For 
large $S$ the particulate eigenvalue tends towards that of the non-particulate case and hence $\lambda_p'\rightarrow 1$.
For $m=0$, in each case there is a clearly defined $S^m$ (marked $\bullet$) for which $\lambda'_p$ is minimised and 
the flow is least stable.
\textbf{Left:} Fixed $\alpha=2$ and $Re=1000$ (line), $3000$ (dashed), $10000$ (short dashed), $20000$ (dots).
There is no qualitative change in behaviour, but the Stokes number for which the effect of the particles changes from 
destabilising to stabilising reduces with $Re$.
\textbf{Right:} Fixed $Re=1000$ and $\alpha=0.2$ (line), $0.4$ (dashed), $1$ (short dashed), $2$ (dots). Again there
is no qualitative change with $\alpha$ but the region of $S$ where the effect changes from destabilising to 
stabilising decreases with $\alpha$.
\label{fig:u_omega_S}}
\end{figure}

In between these two extreme limits, non-trivial changes occur to the leading eigenvalue. For $m=0$ the 
behaviour is readily described. In figure \ref{fig:u_omega_S} (left), $\lambda_p'$ smoothly varies from less stable 
($\lambda_p'\simeq 0.995$ for $S=0$) to unaffected ($\lambda_p'=1$ for $S\rightarrow \infty$), it does not do so 
monotonically. In particular it initially \emph{decreases} the stability of the flow, then \emph{over 
stabilises} the flow past the level of a pure fluid  before it subsides to the particle-free result. 
This occurs for all $Re$ and $\alpha$ considered.

\begin{figure}
\centering
\includegraphics[width=0.8\textwidth]{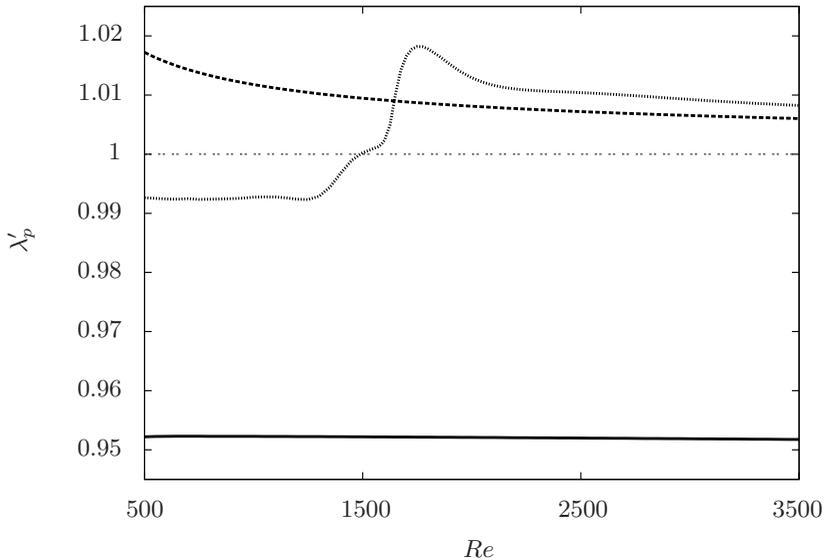}
\caption{Normalised leading growthrate, $\lambda_p'$ for $m=0$ as a function of $Re$, 
for $S=10^{-3}$ (line), $S=10^{-2}$ (dots), $S=10^{-1}$ (dashed) with $f=0.01$, $\alpha=1$. While the largest and 
smallest values of $S$ present straightforward, monotonic behaviour, the intermediate $S=0.01$ presents non-trivial
 variation with Reynolds number.
\label{fig:u_omega_Re}}
\end{figure}

This result is clarified further by fixing $S$ and varying $Re$, as in figure \ref{fig:u_omega_Re}. For low values
of $S$ (here $10^{-3}$) the stability remains effectively unchanged at these Reynolds numbers with the particles 
remaining as passive tracers. For large Stokes number ($S=0.1$) there is some variation of $\lambda_p'$ with $Re$, but 
it is relatively benign as the particles decouple from the flow. It is only at intermediate Stokes number ($S=0.01$) 
that we see nontrivial behaviour for moderate Reynolds number.

The case of $m=1$ is more complex (figure \ref{fig:u_omega_S}, right). The limiting cases of very large or 
very small $S$ still behave as expected (though now even smaller values of $S$ must be considered to recover 
the limiting case) but the intermediate behaviour is more involved. The particles can either stabilise or 
destabilise the flow depending on the precise parameters chosen. For a given $Re$ and $\alpha$, increasing 
$S$ can lead to the flow switching repeated between one and the other.

\begin{figure}
\centering
\includegraphics[width=\textwidth]{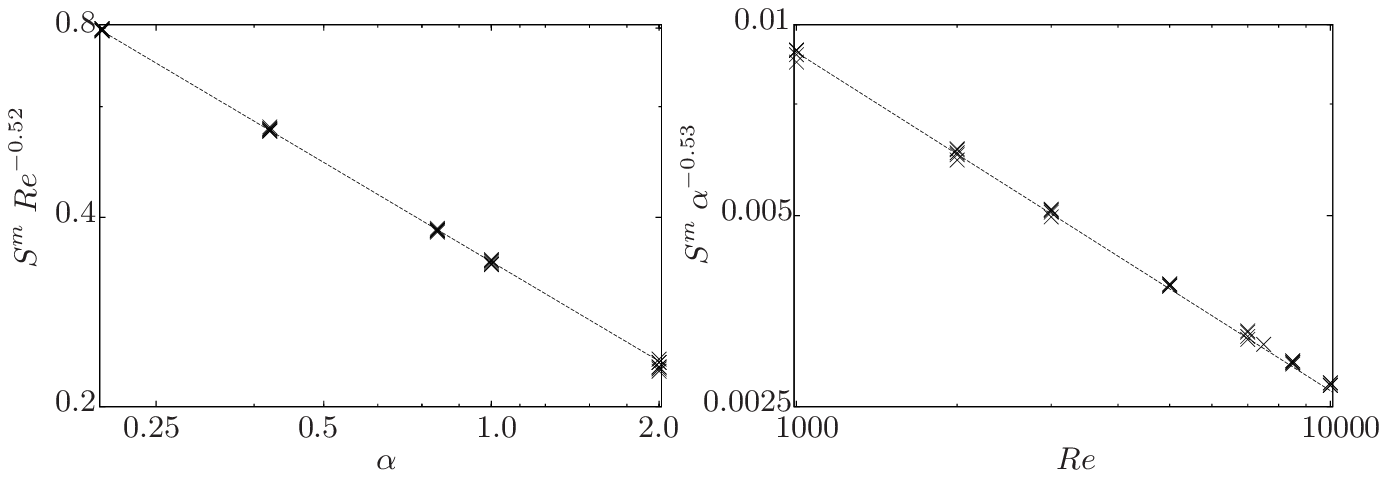}
\caption{The variation $S^m$ with $Re$ and $\alpha$. The data can be collapsed down on to a single line by using an appropriate rescaling.
\textbf{Upper:} $S^mRe^{-0.52}$ (exponent given to two significant figures) as a function of $\alpha$. 
The collapsing of the data onto close to a single line suggests $S^m\propto Re^{0.52}$. 
\textbf{Lower:} $S^m\alpha^{-0.53}$ as a function of $Re$. The data again collapses onto a single line, though not 
as cleanly as for the scaling in $Re$. Nonetheless, this suggests $S^m\propto \alpha^{-0.53}$.}
\label{fig:u_Sm}
\end{figure}

We conclude this section by noting that the simple behaviour for $m=0$ suggests we may be able to isolate some simple
behaviour. To identify the region in which particles have the most significant effect on the flow, we define $S^m$ to be the 
Stokes number for which the flow is most destabilised and $\lambda_p'$ is minimised. Least squares fitting suggests a 
clear scaling of $S_m$ with both $Re$ and $\alpha$ as shown in figure \ref{fig:u_Sm}. While the effect is somewhat 
unsurprisingly amplified at larger Reynolds number, it mostly concerns longer wavelengths and remains limited in 
amplitude in all cases (we never found an increase of growthrate more that 2\% higher than the single fluid case).

%This occurs for all streamwise and azimuthal wavenumbers considered. It is 
%of particular note that although the least stable eigenvalue for both the particulate and non-particulate flows 
%correspond to $m=1$, the azimuthal wavenumber for which the leading eigenvalue is most effected is $m=0$. 

\section{Nonuniform particle distributions}\label{sec:nonuniform}
There is nothing inherent in the model which requires the initial distribution of particles to be uniform. Relaxing 
this assumption allows to us to consider a more general problem, albeit at the cost of including 
%additional 
all terms in the linearised equations (\ref{eqn:lin1})-(\ref{eqn:lin2}). As discussed in the introduction, experimental work 
suggests that for low to moderate Reynolds numbers particles congregate at a particular radius forming an annulus 
from their distribution centred in the region $r=0.5-0.8$. In this section we capture the essence of this 
by considering distributions of the form 
\begin{equation}
N_0(r)=\tilde{N}\exp\{-(r-r_d)^2/2\sigma^2\},
\end{equation}
with $\tilde{N}$ chosen such that $\int_0^1 N(r)rdr=1$.

Throughout this section we keep fix $S=10^{-3}$, $f=0.1$ and $m=1$ to reduce the set of parameters being 
 considered. 
The first two of these is consistent with experimentally realisable parameters (see section \ref{sec:expts}) 
while $m=1$ is the only azimuthal wavenumber for which we observed instability. 

\subsection{The onset of instability}
\begin{figure}
\centering
\includegraphics[width=0.8\textwidth]{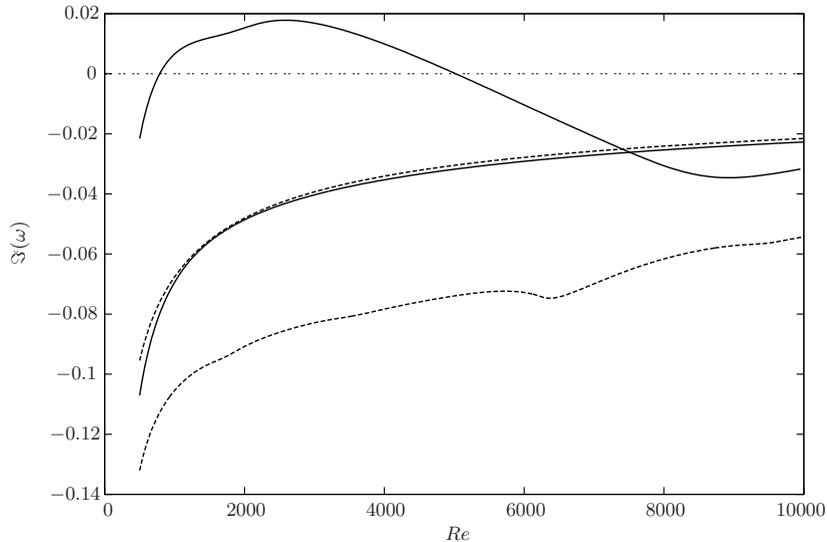}
\caption{The leading eigenvalue for uniform (solid) and non-uniform particle distributions, centred at either 
$r=0.6$ (dashed) or $r=0.7$ (dotted). The uniform distribution is stable for all $Re$, but both the non-uniform distributions are unstable for a range of $Re$. For higher $Re$ the leading eigenmode switches between A and P branches for both non-uniform distributions at the point where dashed and solid lines meet on the graph. }
\label{fig:n_omega_Re}
\end{figure}

As soon as the assumption of uniform particle distribution is relaxed we see a linear instability occurring. Figure 
\ref{fig:n_omega_Re} shows the leading eigenvalues for two different localised distributions of particles compared 
with the uniform distribution result. Whereas the latter of these remains stable for all $Re$, the two non-uniform 
distributions are unstable. Of particular note is that in both cases we see instability for \emph{moderate} 
Reynolds number, but \emph{not} for either high or low $Re$. This initially surprising observation that the 
flow re-stabilises as $Re$ increases is a recurrent observation. For very large $Re$, there is no coupling 
between the fields and everything is stable. For low $Re$, diffusion dominates and imposes stability. 
Only in the middle is instability feasible. 	

\begin{figure}
\centering
\includegraphics[width=0.8\textwidth]{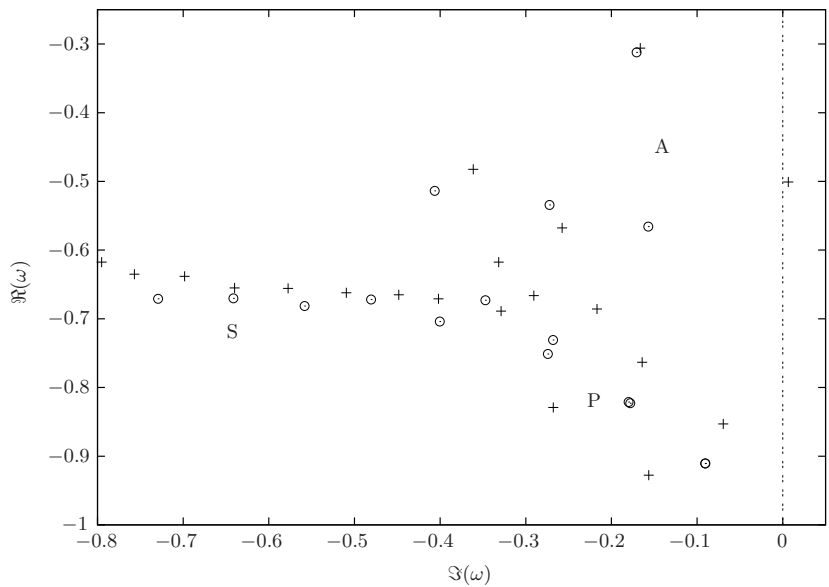}
\caption{Eigenvalue spectra for the non-particulate ($\circ$) and particulate cases ($+$). In both cases 
$Re=1000$, $\alpha=1$ and $m=1$ while the particles were non-uniformly distributed with $f= 0.1$, $r_d=0.6$ 
and $\sigma=0.1$.}\label{fig:n_spec}
\end{figure}

For higher $Re$, after the flow has restabilised we observe there is a switching of the leading eigenmode (at 
around $Re=6000-8000$) after which the dominant eigenmode appears to be the same as for the uniform problem. 
Closer examination of the eigenvalue spectrum (figure \ref{fig:n_spec}) reveals that for an unstable 
configuration, the leading eigenvalue is now in the P-branch of the spectrum, rather than the A-branch as in the case
of both the non-particulate and uniformly distributed problems.

\begin{figure}
\centering
\includegraphics[width=0.8\textwidth]{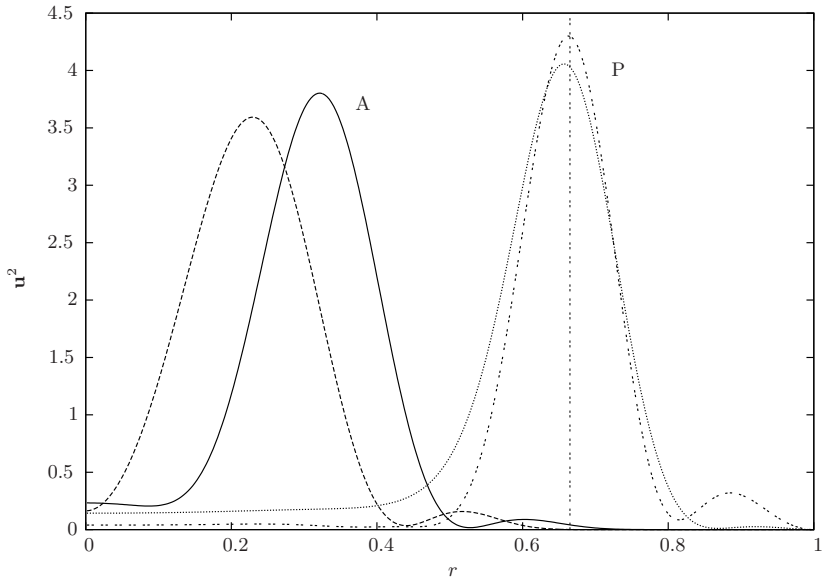}
\caption{The distribution of the fluid energy in the leading eigenmodes. The two that peak on the left are the leading 
A branch modes for the non-particulate (solid line) and particulate case (dashed). 
The leading P modes peak on the right with the non-particulate case being double-dashed and the particulate 
profile dotted. For the particulate case the particles are non-uniformly distributed with $Re=1000$, $m=1$, $f=0.1$, 
$S=10^{-3}$, $r_d=0.7$ and $\sigma=0.1$. The vertical line is at $r_d^*=0.666$.\label{fig:eigenmodes}}
\end{figure}

The reason for the switching of branches becomes clear as soon as we examine the eigenmodes associated with the 
two eigenvalues. In figure \ref{fig:eigenmodes} the leading eigenmodes of the two branches are plotted. The overall 
shape is relatively insensitive to the distribution of particles, but the modes of the two branches are primarily 
active in different parts of the pipe. For the A-branch, the eigenmode is localised to a relatively central 
part of the domain (centred at $r\approx 0.3$), while the P-branch mode is located nearer the edge of the pipe
($r\approx 0.7$). It is unsurprising that when the particle distribution is centred near this outer location, 
these are the eigenmodes that are primarily excited.

\begin{figure}
\centering
\includegraphics[width=0.8\textwidth]{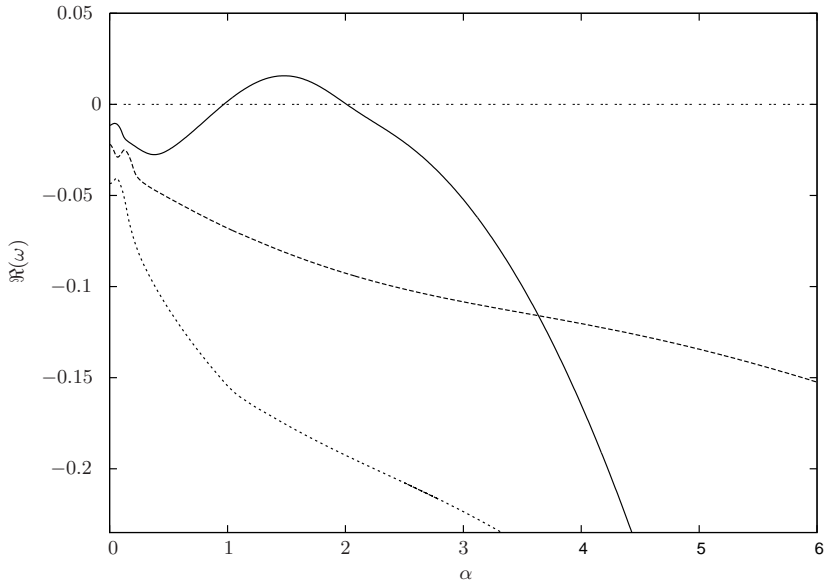}
\caption{The leading growthrates for $Re=1000$, $m=1$, $f=0.1$, $S=10^{-3}$, $r_d=0.65$ and $\sigma=0.1$. 
Instability ($\Im\{\omega\}>0$) only 
occurs for a finite range of $\alpha$, with the flow being stable to both long and short wavelength disturbances. 
\label{fig:omega_alpha}}
\end{figure}

As well as only being unstable for a finite range of $Re$, the flow is also only unstable for a finite range of $\alpha$
(figure \ref{fig:omega_alpha}). For both small and large wavenumber disturbances the flow is stable. The latter is to 
be expected due to the stabilising influence of viscosity, but it is important to note the instability exists at very 
moderate wave numbers for which the model is expected to be valid.

\subsection{Effect of the radial distribution of particles}

\begin{figure}
\centering
\includegraphics[width=0.8\textwidth]{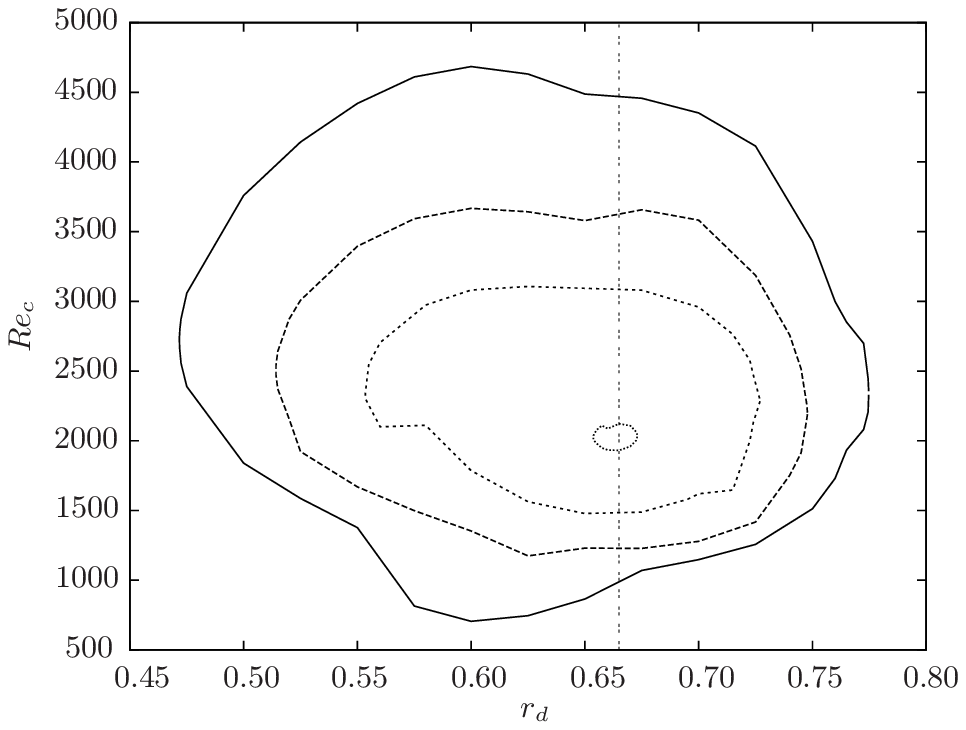}\vspace{3mm}

\includegraphics[width=0.8\textwidth]{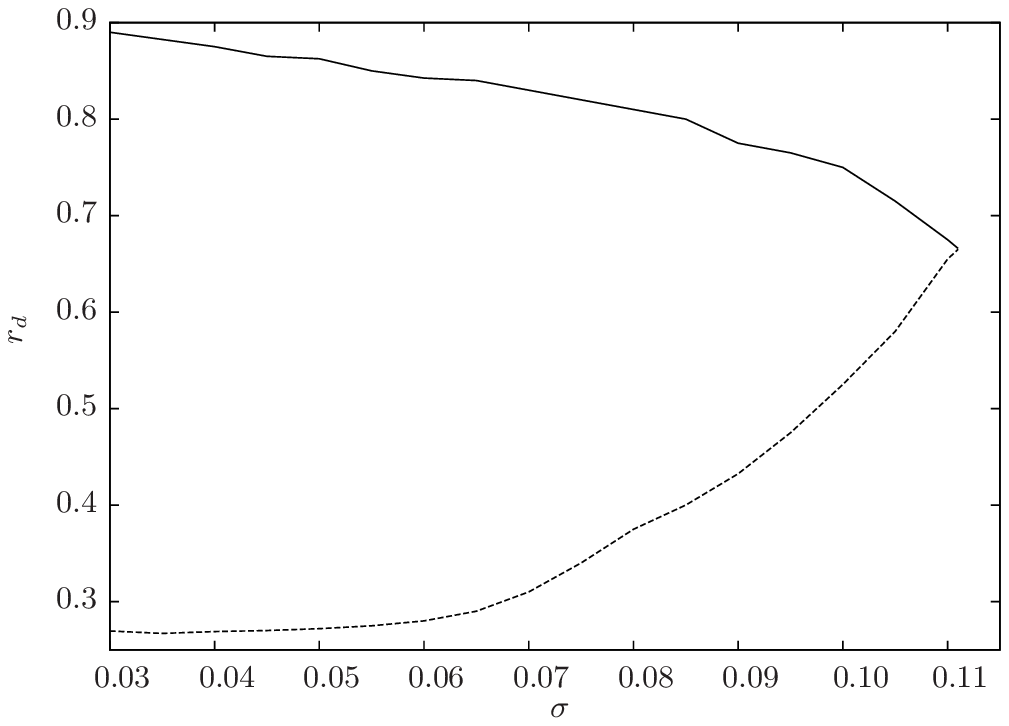}
\caption{\textbf{Upper:} Contours of neutral stability in $Re-r_d$ space for 
values of $\sigma=0.110$, $0.105$, $0.100$ and $0.095$ from innermost to outermost. In each case, the enclosed region 
is the unstable region. That all the contours are closed indicates 
there is a maximum/minimum value of both $Re$ and $r_d$ for which flow is unstable. \textbf{Lower:} The 
maximum (solid)/minimum (dashed) values of $r_d$ for which the flow becomes unstable as $\sigma$ is varied, 
searching across 
all $Re$. There is a maximum value of $\sigma$ beyond which instability isn't possible.}\label{fig:n_rminmax}
\end{figure}

The exact location where the particle annulus ($r_d$) is centred, and how sharply the distribution peaks around this 
location, plays an important role in determining whether the flow 
becomes unstable or not. By searching over $\alpha$ we can trace out neutral stability contours in $Re-r_d$ space  
for differing values $\sigma$ (figure \ref{fig:n_rminmax}, upper). The enclosed regions are unstable and we 
see that all the contours are indeed closed. The fact that there is a minimum/maximum value of $Re$ for which 
the flow is unstable is consistent with our earlier observations, while the fact that there are bounds on the 
value of $r_d$ supports the thesis of needing to excite the P-branch in order to destabilise the flow. We note 
that for all values of $\sigma$ the curves are concentric and the broadest range of unstable $Re$ occurs 
when $r_d$ is in the region $0.6-0.7$.

We track the maximum and minimum values of $r_d$ for which instability exists in figure \ref{fig:n_rminmax} (lower). 
By doing so we arrive at a minimum degree of localisation required to trigger instability, corresponding 
to $\sigma^*=0.111$, for which the particle distribution must be centred at $r_d^*=0.666$.  

\section{Relevance to experimental configurations}\label{sec:expts}
\begin{figure}
\centering
\includegraphics[width=0.95\textwidth]{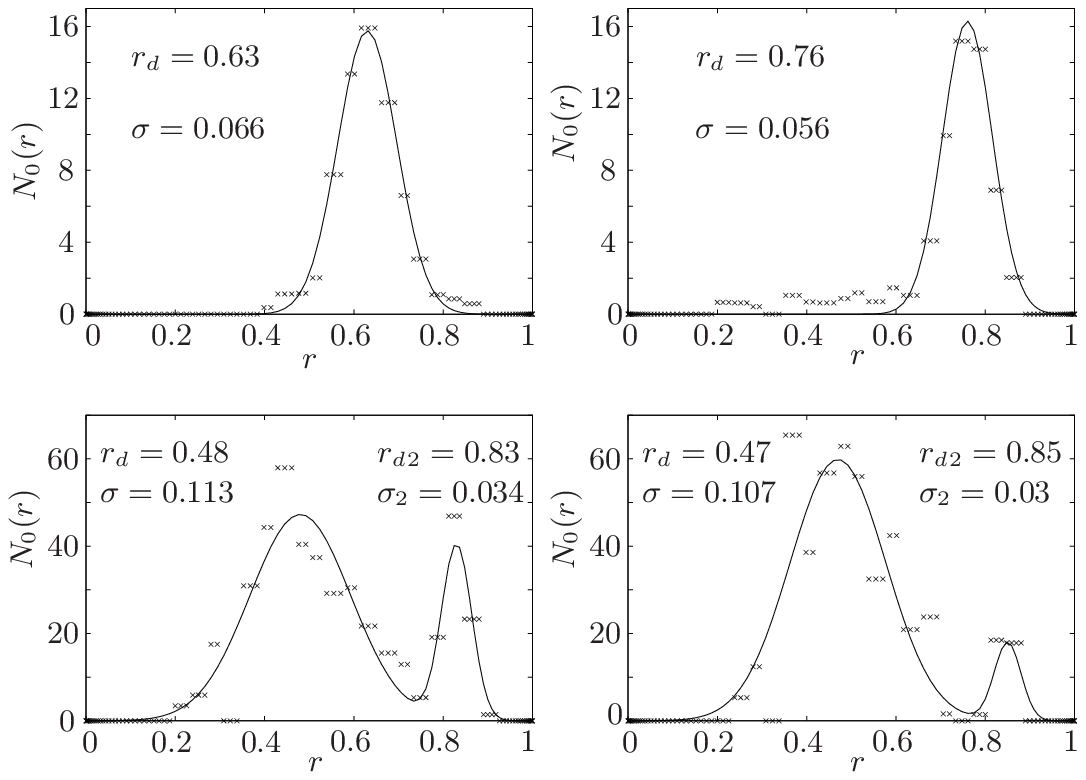}
\caption{Particles concentration as a function of the radius. The crosses show the experimental results of
\citet{matas2004jfm} while the lines are our fitted distributions. For the top row ($Re=67$ (left) and $Re=350$ 
(right)) a single gaussian was fitted for each case, centred at $r_d$ and of width $\sigma$. For the bottom row 
($Re=1000$ (left) and $Re=1650$ (right)) each set of data was fitted with the sum of two gaussians of the given 
locations and widths.
\label{fig:matas}}
\end{figure}	
%
%\begin{table}
%\centering
%\begin{tabular}{| c | c | c | c | c |} 
%\(Re \)& \( S\)  &  \(\omega_0\) (fl)&\(\omega_0\) (gauss)&\(\omega_0\) (matas) \\
%\hline
%  \( 67 \) & \(2.743\times 10^{-3} \)&\( -0.58409   \) &  \(-0.55828 \) &   \(-0.56033 \)   \\[1mm]  
%   \(350\) & \(2.743 \times10^{-3} \) &\( -0.14605  \) &  \(-0.17752 \) &  \(-0.16606\)    \\[1mm]
%  \( 1000\)& \(7.689 \times10^{-4}\) &\( -9.1143\times 10^{-2}  \)  &  \(-0.10635 \) &\( -0.10480 \) \\[1mm]
%  \( 1650 \)& \( 7.689\times 10^{-4}\)  & \( -7.4771 \times10^{-2} \)   & \( -9.9083 \times10^{-2}\)  &\(-9.4478 \times 10^{-2}\)
%\end{tabular}
%\caption{Comparison of leading eigenvalues for the linear stability problem obtained in cases 
%without particles, with particles distributions experimentally found by \cite{matas2004jfm}, and with 
%their closest Gaussian fits. In each case the eigenvalues are given for $m=1$ and $\alpha=1$. \label{table:matas}}
%\end{table}

\begin{table}
\centering
\begin{tabular}{ccccc}
\toprule 
&&\multicolumn{3}{c}{$\Im\{\omega\}$} \\
\cmidrule(r){3-5}
\(Re \)& \( S\)  &  particle free & discontinuous & continuous \\
\midrule
  \( 67 \) & \(2.743\times 10^{-3} \)&\( -0.58409   \)  &   \(-0.56033 \) &  \(-0.55828 \)  \\[1mm]  
   \(350\) & \(2.743 \times10^{-3} \) &\( -0.14605  \)  &  \(-0.16606\)  &  \(-0.17752 \)  \\[1mm]
  \( 1000\)& \(7.689 \times10^{-4}\) &\( -0.091143  \)   &\( -0.10480 \) &  \(-0.10635 \)\\[1mm]
  \( 1650 \)& \( 7.689\times 10^{-4}\)  & \( -0.074771 \)   &\(-0.094478 \) & \( -0.099083 \)  \\
  \bottomrule
\end{tabular}
\caption{Comparison of leading eigenvalues for the linear stability problem obtained in the cases of no particles, 
with particles distributions experimentally found by \cite{matas2004jfm} (see figure \ref{fig:matas}), and with 
particles distributed with closest Gaussian fits to the experimental data (the parameters given in figure 
\ref{fig:matas}). In each case the eigenvalues are given for $m=1$ and $\alpha=1$. \label{table:matas}}
\end{table}

\citet{matas2004jfm} explores the effect of adding particles to pipe flow. As with other experimental work they 
report the clustering of particles at preferential radii that motivates this study but they do not report evidence 
of a linear instability. In this section we analyse the configurations observed by \citeauthor{matas2004jfm} and 
show that our numerical results are consistent with the experiments - that is that we find the configurations to be 
linearly stable.

In the experimental work, four configurations of particles are explicitly given (figure \ref{fig:matas}) corresponding 
to $Re=67$, $350$, $1000$ and $1650$ (left to right, top to bottom). 
At low $Re$ the particles all cluster at a single radius consistent with 
\citet{segre1962jfm}. As the $Re$ is increased, two preferential radii emerge and coexist. We capture these 
distributions within the linear stability analysis with two approaches. Firstly we fit either one or two Gaussian 
distributions through the data using least squares. These fits and the corresponding fitting parameters are those 
given in \ref{fig:matas}. Secondly, we use the raw data to give a discontinuous distribution with the $N(r)$ being taken 
as constant between data points. 

In table \ref{table:matas} we give the growth rates of the leading eigenvalues for non-particulate flow, 
particles distributed continuously and particles distributed discontinuously for the 
different configurations reported by \citeauthor{matas2004jfm}. For $Re=67$ both distributions of particles reduce the
stability of the flow, but not so far as to make it unstable. For the higher values of $Re$, the particles in fact 
stabilise the flow further. These effects apply for both the Gaussian and discontinuous particle distributions and all 
growth rates agree to within at most $7\%$, much less than the discrepancy with the non-particulate case. We conclude 
that within the set of cases experimentally studied, our numerical results are fully consistent with the observations.

\section{Conclusions and discussion}\label{sec:conc}
We have presented a very simple model for particulate pipe flow. Although this model has been examined before 
in plane shear flow \citep{saffman62,klinkenberg11}, it has only been done in the context of an initially uniform 
particle distribution. In that previous work, the flow has always remained stable and here this is observed for pipe flow too. 
We are able to track the curves in parameter space for which the flow becomes most effected by the presence of 
particles, but it does always remain stable.

Relaxing the assumption of uniformly distributed particles, and allowing for the experimentally 
observed situation of particles arranged preferentially in an annulus is sufficient to induce linear 
instability in the flow for certain ranges of parameter. In particular, the flow is only ever unstable for 
intermediate Reynolds numbers, restabilising as $Re$ is increased further. This in-between regime is sandwiched 
between low $Re$ flows dominated by viscous diffusion and high $Re$ flows where the two phase decouple. 
The instability also only exists at intermediate axial wavenumbers. This avoids both the small length scale 
disturbances which violate the assumptions of the model and also the large (axial) scale disturbances which 
must test any assumption of axial independence of the base state.

The linear instability appears strongest when the annulus of particles is centred at $r_d\approx 0.65$ both in terms 
of this being the location where the smallest degree of localisation is needed for instability and being 
closely correlated with the widest band of unstable $Re$ for stronger localisation. This is particularly important 
as experimental work suggests that particles naturally congregate at this radius. The experimental work done 
to date on transitional particulate pipe flow \citep{matas2003prl,matas2004jfm} has all been within 
the region of parameter space that this study has found to be linearly stable and so is entirely consistent with this. 

That linear instability is feasible even within such a simple framework highlights the complexities of the problem 
and reveals that very different transition scenarios can be at play within the broader problem of particulate pipe 
flow. We do not submit this as a full explanation for the transition problem not least because it is possible 
that some of the excluded physics has a stabilising effect on the flow. In particular, the inertial mechanisms 
driving the particles to form into an annulus could be expected to act as a 
stabilising influence. Instead we suggest that the formation of an annulus of particles could be a key step in 
the onset of turbulence for certain, experimentally relevant parametric configurations of the problem.

\section{Acknowledgements}
AR is supported by TUV-NEL. 
CCTP is partially supported by EPSRC grant No. EP/P021352/1 .
AP acknowledges support from the Royal Society under the Wolfson Research Merit Award Scheme (Grant WM140032). 
We thank Ashley Willis for use his code as well as useful discussions on adapting it.

\bibliographystyle{apalike}
\bibliography{particulate_bib}
\end{document}